\documentclass{PoS}

\usepackage{epsf}
\usepackage{amssymb}
\usepackage{amsmath}
\usepackage{amsfonts}
\usepackage{cite}
\usepackage[small]{caption2}

\usepackage[T1]{fontenc}
\usepackage{psfrag,epsfig,graphicx,graphics}

\newcommand{\be}{\begin{equation}}
\newcommand{\beq}{\begin{equation}}
\newcommand{\ee}{\end{equation}}
\newcommand{\eq}{\end{equation}}
\newcommand{\eeq}{\end{equation}}

\newcommand{\bea}{\begin{eqnarray}}
\newcommand{\eea}{\end{eqnarray}}

\def\slashchar#1{\setbox0=\hbox{$#1$}
   \dimen0=\wd0
   \setbox1=\hbox{/} \dimen1=\wd1
   \ifdim\dimen0>\dimen1
      \rlap{\hbox to \dimen0{\hfil/\hfil}}
      #1
   \else
      \rlap{\hbox to \dimen1{\hfil$#1$\hfil}}
      /gdatdafinal2.tex
   \fi}

\newcommand{\veck}{{\bf k}}

\newcommand{\veckone}{{\bf k}_1}
\newcommand{\vecktwo}{{\bf k}_2}
\newcommand{\veckj}{{\bf k}_{J}}

\newcommand{\veckjone}{{\bf k}_{J,1}}
\newcommand{\veckjtwo}{{\bf k}_{J,2}}

\newcommand{\deins}[1]{{\rm d}#1\,}
\newcommand{\dzwei}[1]{{\rm d}^2#1\,}
\newcommand{\dk}{\dzwei{\veck}}

\newcommand{\dkone}{\dzwei{\veckone}}
\newcommand{\dktwo}{\dzwei{\vecktwo}}

\newcommand{\dsigma}{\deins{\sigma}}
\newcommand{\dsigmahat}{\deins{{\hat\sigma}_{\rm{ab}}}}
\newcommand{\dnu}{\deins{\nu}}

\newcommand{\dx}{\deins{x}}
\newcommand{\dxone}{\deins{x_1}}
\newcommand{\dxtwo}{\deins{x_2}}
\newcommand{\dyjetone}{\deins{y_{J,1}}}
\newcommand{\dyjettwo}{\deins{y_{J,2}}}
\newcommand{\dphij}{\deins{\phi_{J}}}
\newcommand{\dphijone}{\deins{\phi_{J,1}}}
\newcommand{\dphijtwo}{\deins{\phi_{J,2}}}
\newcommand{\dtwojets}{{\rm d}|\veckjone|\,{\rm d}|\veckjtwo|\,\dyjetone \dyjettwo}
\newcommand{\shat}{{\hat s}}

\newcommand{\asbar}{{\bar{\alpha}}_s}

\newcommand{\chihat}{{\omega}}

\title{First complete NLL BFKL calculation of Mueller Navelet jets at LHC}

\ShortTitle{First complete NLL BFKL calculation of Mueller Navelet jets at LHC}

\author{D.~Colferai\\
        Dipartimento di Fisica, Universit{\`a} di Firenze, Italy\\
        INFN, Florence, Italy\\
        Email: \email{colferai@fi.infn.it}}

\author{F.~Schwennsen\\
        Deusches Elektronen-Synchrotron DESY, Hamburg, Germany \\
        Email: \email{florian.schwennsen@desy.de}}

\author{L.~Szymanowski\\
        Soltan Institute for Nuclear Studies, Warsaw, Poland {\em \&} \\
        CPHT, {\'E}cole Polytechnique, CNRS, 91128 Palaiseau Cedex, France\\ 
        Email: \email{lech.szymanowski@fuw.edu.pl}}

\author{\speaker{S.~Wallon}
\\
LPT, Universit{\'e} Paris-Sud, CNRS, 91405 Orsay, France \ {\em \&} \\
UPMC Univ. Paris 06, facult\'e de physique, 4 place Jussieu, 75252 Paris Cedex  05, France\\
        E-mail: \email{wallon@th.u-psud.fr}}

\abstract{For the first time, a next-to-leading BFKL study of the cross section and azimuthal decorrellation of Mueller Navelet jets is performed, i.e. including next-to-leading corrections to the Green's function as well as next-to-leading corrections to the Mueller Navelet vertices. The obtained results for standard observables proposed for studies of Mueller Navelet jets show that both sources of corrections are of equal and big importance for final magnitude and final behavior of observables, in particular for the LHC kinematics investigated here in detail. The astonishing conclusion of our analysis is that the observables obtained within the complete next-lo-leading order BFKL framework of the present paper are quite similar to the same observables obtained within next-to-leading logarithm DGLAP type treatment. The only noticeable difference is the ratio the azimuthal angular moments  $\langle \cos 2\varphi\rangle / \langle \cos \varphi\rangle$ which still differs in both treatments.
}

\FullConference{35th International Conference of High Energy Physics\\
                 July 22-28, 2010\\
                 Paris, France}

\begin{document}

\section{Introduction}
\label{Sec_Int}

The high energy regime of QCD is one the key questions of particle physics.
In the semi-hard regime of a scattering process in which $s \gg -t$,  logarithms of the type $[\alpha_s\ln(s/|t|)]^n$ have to be resummed, giving the 
%
%
%
leading logarithmic (LL) Balitsky-Fadin-Kuraev-Lipatov (BFKL) \cite{BFKL_LL} Pomeron contribution to the gluon Green's function describing the exchange in the $t$-channel.
The question of testing such effects experimentally then appeared. Based on new experimental facilities, characterized by increasing center-of-mass energies and luminosities, various  observables have been proposed and often tested in inclusive \cite{test_inclusive}, semi-inclusive  \cite{test_semi_inclusive} and exclusive processes \cite{test_exclusive}. The basic idea is to select specific observables in order to minimize  standard collinear logarithmic effects \`a la DGLAP \cite{DGLAP}  with respect to the BFKL one.
This aims to choose them in such a way that the involved transverse scales would be of similar order of magnitude.
%

We will here consider one of the 
most famous testing ground for BFKL physics: the Mueller Navelet jets \cite{Mueller:1986ey} in hadron-hadron colliders, defined as being separated by a large relative rapidity, while having two similar transverse energies. One thus expects an almost back-to-back emission in a DGLAP scenario, while the allowed emission of partons between these two jets in a BFKL treatment leads in principle to a larger cross-section, without azimuthal correlation between them.
The predicted power like rise of the cross section with increasing energy has been observed at the Tevatron $p\bar{p}$-collider \cite{Abbott:1999ai}, but the measurements revealed an even stronger rise than predicted by BFKL calculations.  Besides, the leading logarithmic approximation \cite{DelDuca:Stirling} overestimates this decorrelation by far. Improvements have been obtained by taking into account some corrections of higher order like the running of the coupling \cite{Orr:Kwiecinski}. Calculation with the full NLL BFKL Green's function
\cite{BFKL_NLL} have been published recently \cite{Vera:Marquet}.
We here review results of Ref.~\cite{us_MN} on
the full NLL BFKL calculation where also the NLL result for the Mueller Navelet vertices \cite{Bartels:vertex} is taken into account. 

\section{NLL calculation}
\label{sec:NLLcalculation}


The two hadrons collide at a center of mass energy $s$ producing two very forward jets, whose transverse momenta  are labeled by Euclidean two dimensional vectors $\veckjone$ and $\veckjtwo$, while their azimuthal angles are noted as $\phi_{J,1}$ and $\phi_{J,2}$. We will denote the rapidities of the jets by $y_{J,1}$ and $y_{J,2}$ which are related to the longitudinal momentum fractions of the jets via $x_J = \frac{|\veckj|}{\sqrt{s}}e^{y_J}$.
%
%
Since at the LHC the binning in rapidity and in transverse momentum will be quite narrow \cite{Cerci:2008xv}, we consider the case of fixed rapidities and transverse momenta. 
%
%
Due to the large longitudinal momentum fractions $x_{J,1}$ and $x_{J,2}$
of the forward jets, collinear factorization holds and the differential cross section can be written as
\begin{equation}
  \frac{\dsigma}{\dtwojets} = \sum_{{\rm a},{\rm b}} \int_0^1 \dxone \int_0^1 \dxtwo f_{\rm a}(x_1) f_{\rm b}(x_2) \frac{\dsigmahat}{\dtwojets},
\end{equation}
where $f_{\rm a,b}$ are the parton distribution functions~(PDFs) of a parton a (b) in the according proton (which are renormalization scale $\mu_R$ and  factorization scale $\mu_F$ dependent).
%
The  resummation of logarithmically enhanced contributions 
are included 
through $k_T$-factorization:
\begin{equation}
  \frac{\dsigmahat}{\dtwojets} = \int \dphijone\dphijtwo\int\dkone\dktwo V_{\rm a}(-\veckone,x_1)G(\veckone,\vecktwo,\shat)V_{\rm b}(\vecktwo,x_2),\label{eq:bfklpartonic}
\end{equation}
where the BFKL Green's function $G$ depends on $\shat=x_1 x_2 s$. The jet vertices
 $V_{a,b}$ were  calculated at
 NLL order in Ref.~\cite{Bartels:vertex}.
%
Combining the PDFs with the jet vertices one writes
\begin{eqnarray}
  \frac{\dsigma}{\dtwojets} 
&=& \int \dphijone\dphijtwo\int\dkone\dktwo \Phi(\veckjone,x_{J,1},-\veckone)\,G(\veckone,\vecktwo,\shat)\,\Phi(\veckjtwo,x_{J,2},\vecktwo) \,\nonumber
\\
\mbox{where }&& \quad \Phi(\veckjtwo,x_{J,2},\vecktwo) = \int \dxtwo f(x_2) V(\vecktwo,x_2).
\end{eqnarray}
%
%
In view of the azimuthal decorrelation we want to investigate, we define the following coefficients:
\beq
  \mathcal{C}_m \equiv \!\!\!\int \!\! \dphijone\dphijtwo\cos\big(m(\phi_{J,1}-\phi_{J,2}-\pi)\big)\!\!\!\int\!\!\dkone\dktwo \Phi(\veckjone,x_{J,1},-\veckone)G(\veckone,\vecktwo,\shat)\Phi(\veckjtwo,x_{J,2},\vecktwo) , \nonumber
\eq
from which one can easily obtain the differential cross section and azimuthal decorrelation as
\begin{equation}
  \frac{\dsigma}{\dtwojets} = \mathcal{C}_0 \quad {\rm and} \quad 
  \langle\cos(m\varphi)\rangle \equiv \langle\cos\big(m(\phi_{J,1}-\phi_{J,2}-\pi)\big)\rangle = \frac{\mathcal{C}_m}{\mathcal{C}_0} .
\end{equation}
The guiding principle is then to rely
on the LL-BFKL eigenfunctions
\begin{equation}
  E_{n,\nu}(\veckone) = \frac{1}{\pi\sqrt{2}}\left(\veckone^2\right)^{i\nu-\frac{1}{2}}e^{in\phi_1}\,,
\label{def:eigenfunction}
\end{equation}
although they
 strictly speaking do not diagonalize the NLL BFKL kernel.
In the LL approximation, 
\begin{equation}
  \mathcal{C}_m = (4-3\delta_{m,0})\int \dnu C_{m,\nu}(|\veckjone|,x_{J,1})C^*_{m,\nu}(|\veckjtwo|,x_{J,2})\left(\frac{\shat}{s_0}\right)^{\chihat(m,\nu)}\,,
\label{eq:cm2}
\end{equation}
where
\begin{align}
  C_{m,\nu}(|\veckj|,x_{J})
=& \int\dphij\dk \dx f(x) V(\veck,x)E_{m,\nu}(\veck)\cos(m\phi_J) \,, \label{eq:mastercnnu}
\end{align}
and
$
  \chihat(n,\nu) = \asbar\chi_0\left(|n|,\frac{1}{2}+i\nu\right) ,$ with
$\chi_0(n,\gamma) = 2\Psi(1)-\Psi\left(\gamma+\frac{n}{2}\right)-\Psi\left(1-\gamma+\frac{n}{2}\right)
$
and $\asbar = N_c\alpha_s/\pi$. 
%
%
The master formulae of the LL calculation~(\ref{eq:cm2}, \ref{eq:mastercnnu}) will also be used for the NLL calculation. The price to pay in order that
 $E_{n,\nu}$ remains eigenfunction is to accept that the eigenvalue become an operator containing a derivative with respect to $\nu$. 
In combination with the impact factors the derivative acts on the impact 
factors and effectively leads to a contribution to the eigenvalue
 which depends on the impact factors. 



At NLL, the jet vertices are intimately dependent on the jet algorithm ~\cite{Bartels:vertex}. We here use  the cone algorithm, which is expected to the used by the CMS collaboration. At NLL,
one should also pay attention to the choice of scale $s_0$. We find the choice of
scale  $s_0 =\sqrt{s_{0,1} \, s_{0,2}}$  with 
$s_{0,1}= \frac{x_{1}^2}{x_{J,1}^2}\veckjone^2$ rather natural, since it does not depend on the momenta $\veck_{1,2}$ to be integrated out. Besides, the dependence with respect to $s_0$ of the whole amplitude can be studied,
when taking account the fact that both the NLL BFKL Green function and the vertex functions are $s_0$ dependent.
In order to study the effect of possible collinear improvement \cite{Salam:1998tj,resummed},
we have, in a separate study, implemented for $n=0$ the scheme 3 of Ref.~\cite{Salam:1998tj}. This is only required by the Green function since we could show by a numerical study that the 
jet vertices are free of $\gamma$ poles and thus do not call for any collinear improvement.
In practice, the use of Eqs.~(\ref{eq:cm2}, \ref{eq:mastercnnu}) leads 
to the possibility to calculate for a limited number of $m$ the coefficients $C_{m,\nu}$ as universal grids in $\nu$, instead of using a two-dimensional grid in $\veck$ space.
We use MSTW 2008 PDFs \cite{Martin:2009iq} and a two-loop strong coupling with a scale $\mu_R= \sqrt{|\veckjone|\cdot |\veckjtwo|}\,.$ Although a BFKL treatment does not require
in principle any asymmetry between the two emmited jets, large logarithms of non-BFKL origin,
of Sudakov type, could dominate the observable for very symmetric jet configurations.
In order 
to compare our analysis with DGLAP NLO approaches \cite{Fontannaz} obtained through the NLL-DGLAP partonic generator \textsc{Dijet} \cite{Aurenche:2008dn}, for which symmetric configurations lead to instabilities, we here display our results for $|\veckjone|=35\,{\rm GeV}$, $|\veckjtwo|=50\,{\rm GeV}$, and compare them with the DGLAP NLO prediction (see Ref.~\cite{us_MN} for symmetric configurations).

\section{Results}
\label{sec:Results}

\begin{figure}[h!]
  \centering
  \psfrag{varied}{}
  \psfrag{cubaerror}{}
  \psfrag{C0}{\raisebox{.1cm}{\footnotesize $\mathcal{C}_0 \left[\frac{\rm nb}{{\rm GeV}^2}\right] = \sigma$}}
  \psfrag{Y}{\footnotesize$Y$}
 \hspace{-.2cm} \includegraphics[width=4.7cm]{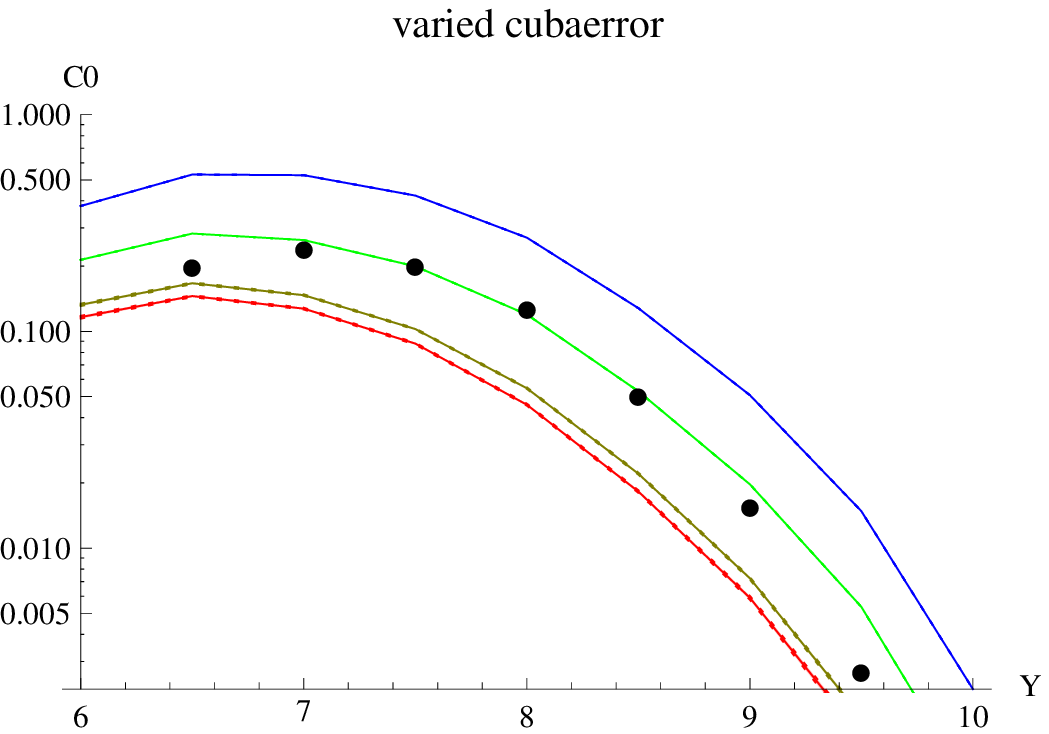} \quad
   \psfrag{C1C0}{\raisebox{.1cm}{\footnotesize $\frac{\mathcal{C}_1}{\mathcal{C}_0}=\langle \cos \varphi\rangle$}}
\includegraphics[width=4.7cm]{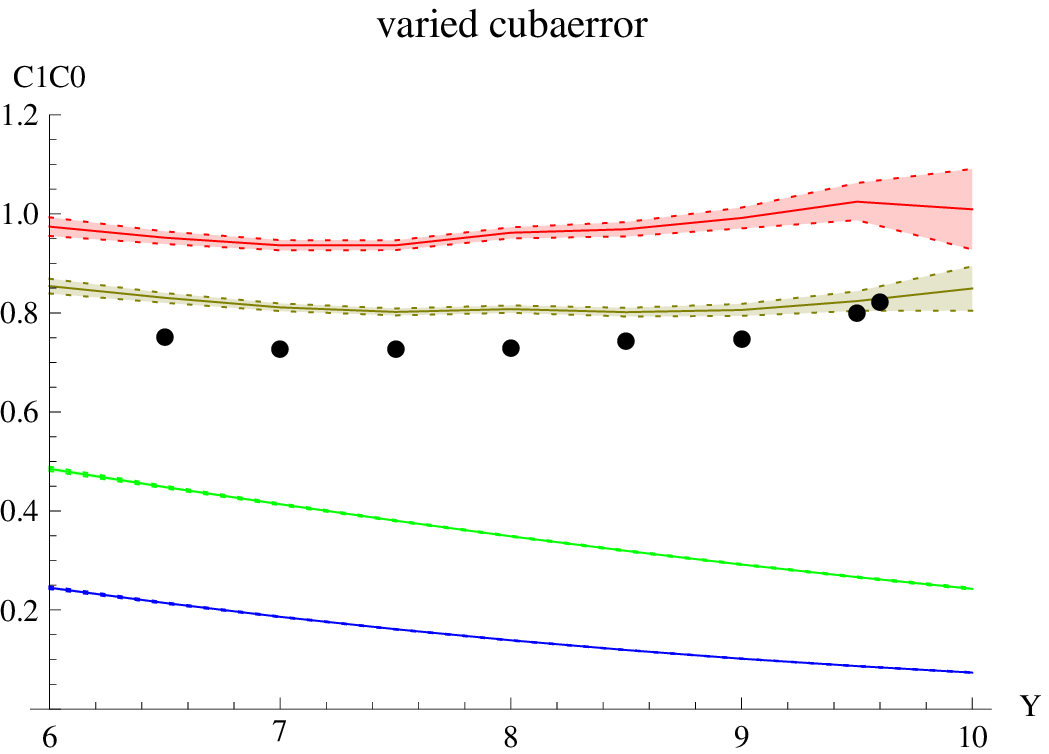}\quad 
 \psfrag{varied}{}
   \psfrag{cubaerror}{}
   \psfrag{C2C1}{\raisebox{.1cm}{\footnotesize$\frac{\mathcal{C}_2}{\mathcal{C}_1}=\langle \cos 2\varphi\rangle / \langle \cos \varphi\rangle$}}
   \psfrag{Y}{\footnotesize$Y$}
  \includegraphics[width=4.7cm]{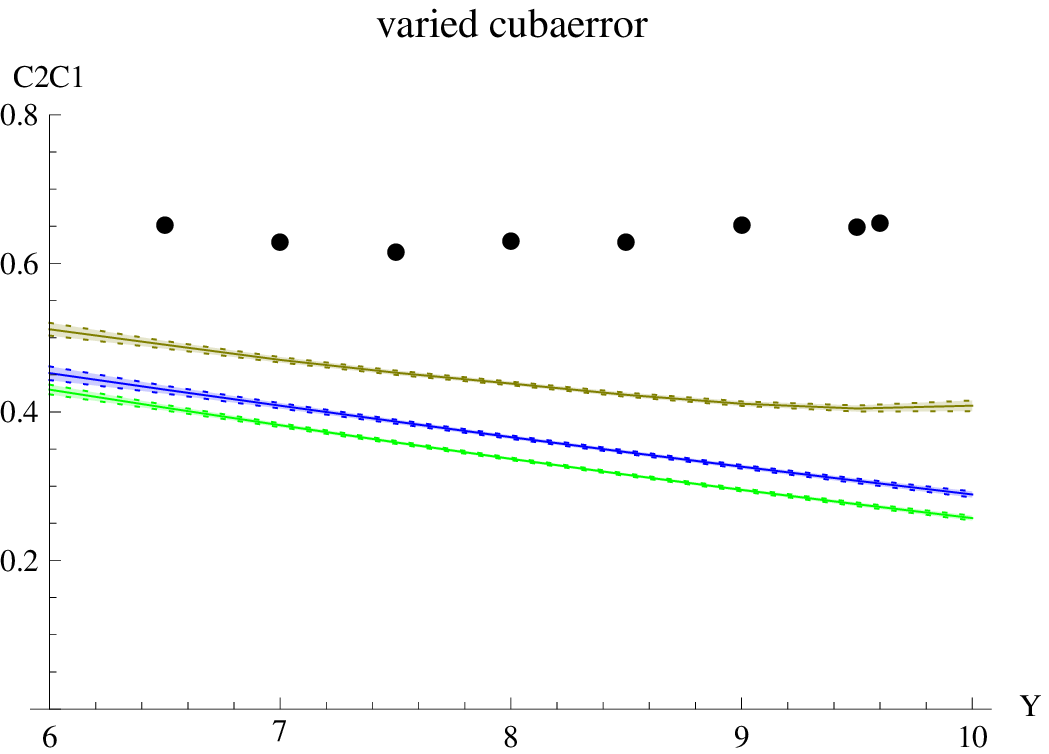}
  \caption{Differential cross section (a), azimuthal correlation $\langle \cos \varphi\rangle$ (b) and ratio $\langle \cos 2\varphi\rangle / \langle \cos \varphi\rangle$ (c) in dependence on $Y$ for $|\veckjone|=35\,{\rm GeV}$, $|\veckjtwo|=50\,{\rm GeV}$. The errors due to the Monte Carlo integration  are given as error bands. 
  As dots are shown the results of Ref.~\cite{Fontannaz} obtained with \textsc{Dijet} \cite{Aurenche:2008dn}.}
  \label{fig:c03550_c1c03550}
\end{figure}

We now present our results for the LHC at the design center of mass energy $\sqrt{s}=14\,{\rm TeV}$. Motivated by a recent CMS study \cite{Cerci:2008xv} we restrict the rapidities of the Mueller Navelet jets to the region $3<|y_J|<5$, thus limiting the relative rapidity $Y$ between 6 and 10 units. Fig.~\ref{fig:c03550_c1c03550}a and ~\ref{fig:c03550_c1c03550}b  respectively
display the cross-section and the azimuthal correlation\footnote{We use the same color coding for both plots, namely blue shows the pure LL result, brown the pure NLL result, green the combination of LL vertices with the collinear improved NLL Green's function, red the full NLO vertices with the collinear improved NLL Green's function.}.
This explicitely shows 
the dramatic effect of the NLL vertex corrections, of the same order as the one for the Green function. In particular, the decorrelation based on our full NLL analysis is very small, similar to the one based on NLO DGLAP.
%
%
The main source of uncertainties is due to the renormalization scale $\mu_R$ and to the energy scale $\sqrt{s_0}$. \ This is particularly important for the azimuthal correlation, which, 
 when including a collinear improved Green's function, may exceed 1 for small $\mu_R=\mu_F$.
The only remaining observable for which 
a noticeable difference can be expected between BFKL and DGLAP type of treatment is the ratio 
$\langle \cos 2\varphi\rangle / \langle \cos \varphi\rangle\,,$ which would deserve a precise experimental analysis.

\vspace{.2cm}

%

%
%
%
%

In conclusion, contrarily to the general belief, the study of Mueller Navelet jets may not be conclusive to exhibit differences between BFKL and DGLAP dynamics.
\vspace{.1cm}

Work supported in part by the Polish Grant N202 249235, the grant ANR-06-JCJC-0084 and by a PRIN grant (MIUR, Italy).

\end{document}